# From Artifacts to Aggregations: Modeling Scientific Life Cycles on the Semantic Web

Alberto Pepe[1,a], Matthew Mayernik[1], Christine L. Borgman[1], Herbert Van de Sompel[2]

1. Center for Embedded Networked Sensing and
Department of Information Studies
University of California, Los Angeles
Los Angeles, CA 90095

2. Digital Library Research and Prototyping Team
Los Alamos National Laboratory
Los Alamos, NM 87545

a. Corresponding author: Alberto Pepe, *apepe@ucla.edu*

Table of Contents

# ABSTRACT

In the process of scientific research, many information objects are generated, all of which may remain valuable indefinitely. However, artifacts such as instrument data and associated calibration information may have little value in isolation; their meaning is derived from their relationships to each other. Individual artifacts are best represented as components of a life cycle that is specific to a scientific research domain or project. Current cataloging practices do not describe objects at a sufficient level of granularity nor do they offer the globally persistent identifiers necessary to discover and manage scholarly products with World Wide Web standards. The Open Archives Initiative's Object Reuse and Exchange data model (OAI-ORE) meets these requirements. We demonstrate a conceptual implementation of OAI-ORE to represent the scientific life cycles of embedded networked sensor applications in seismology and environmental sciences. By establishing relationships between publications, data, and contextual research information, we illustrate how to obtain a richer and more realistic view of scientific practices. That view can facilitate new forms of scientific research and learning. Our analysis is framed by studies of scientific practices in a large, multi-disciplinary, multi-university science and engineering research center, the Center for Embedded Networked Sensing (CENS).

# INTRODUCTION

Over the last century, scientific communication has relied heavily on the dissemination of papers, journal articles, and monographs. Libraries had well established cataloging and access mechanisms to support these products when all were in print form. Librarians devoted most of their cataloging efforts to monographs and journals, leaving the description of individual journal articles and papers to abstracting and indexing services. Now scientific artifacts originate in digital form and exist in a much wider array of genres. These include manuscripts, publications, data, laboratory and field notes, instrument calibrations, preprints, grant proposals, talks, slides, patent applications, theses, dissertations, and genres specific to individual disciplines. Neither scientists nor librarians are coping well with this deluge (Bell, Hey & Szalay, 2009; Borgman, 2007; Hey & Hey, 2006; Hey & Trefethen, 2005).

The processes by which science is conducted have also remained remarkably stable, if only at the most general level. Kenneth Mees, writing in *Nature* almost a century ago, identified three stages of scientific knowledge production: "first, the production of new knowledge by means of laboratory research; secondly, the publication of this knowledge in the form of papers and abstracts of papers; thirdly, the digestion of the new knowledge and its absorption into the general mass of information by critical comparison with other experiments on the same or similar subjects." (Mees, 1918: 355). Subsequent studies of scholarly communication have affirmed this general sequence of research activities for the sciences, whether or not based in the laboratory, and for many other empirical disciplines (Latour, 1987; Meadows, 1974; 1998).

How are we to "digest and absorb" the deluge of digital artifacts into today's "general mass of information"? "Critical comparison" is as essential today as in Mees's day. Given the sophisticated tools now available, comparison should become easier. However, a wider range of scholarly artifacts is now publicly available, and those artifacts often exist in multiple versions or multiple stages of development. New tools, services, and practices are needed to facilitate the management of scholarly products, and to do so in ways that remain true to scholarly processes. These tools must support searching and access both by humans, who can make judgments about relationships, and by computer programs that can follow links representing relationships. A major challenge for the next generation of cyberinfrastructure is to enable discovery and exchange of these disparate scholarly materials and their relationships, in order to facilitate new forms of scholarship and learning (Cyberinfrastructure Vision for 21st Century Discovery, 2007; Borgman, 2007; Van de Sompel & Lagoze, 2007).

In this article, we demonstrate the value of aggregating scholarly resources for discovery with an empirical study conducted in a large National Science Foundation Science and Technology Center, the Center for Embedded Networked Sensing (CENS). Several years of research on scientific data practices in this center have enabled us to identify the artifacts produced and relationships among them. The Open Archives Initiative's Object Reuse and Exchange (OAI-ORE) data model offers a means to represent those artifacts and relationships formally. In this article, we present a conceptual implementation of OAI-ORE in two case studies of embedded sensing research: a seismological project and an environmental sensing project of CENS. We demonstrate that, once represented in the OAI-ORE framework, these artifacts can be discovered, accessed, exchanged, and referenced as components of the same conceptual aggregation, despite being distributed across the Web (Object Reuse and Exchange, 2009; Open Archives Initiative, 2009).

## RATIONALE

The use of identifiers and descriptors to manage and to discover scholarly artifacts is hardly a new idea in print or in digital environments. What is new in the context of data- and information-intensive, distributed, collaborative, multi-disciplinary research enabled by cyberinfrastructure – hereafter referred to as eResearch – is the granularity of description and the scale of artifacts to be described. In a print world, globally unique identifiers are assigned only at the level of whole books (International Standard Book Number, 2009) and whole journals (International Standard Serial Number, 2009). These identifiers, in combination with bibliographic descriptions (some of which express relationships; for example, name changes of journals) allow publishers and libraries jointly to handle acquisition and management.

Lacking unique identifiers for finer granularities such as individual journal articles and papers, discovery depends on descriptions such as bibliographic citations and records from abstracting and indexing services. Descriptions alone result in ambiguities and unreliable retrieval; citations in publications to articles, papers, books, and Web pages are notoriously inconsistent. Cataloging rules, in combination with their digital representations (e.g., the Anglo-American Cataloging Rules and the U.S. MARC format (MARC Standards, 2009)) are able to express properties of an object (e.g., MARC tag 130: Uniform Title) and relationships of an object to other objects (e.g., MARC tag 765: Original Language Entry). However, cataloging rules and MARC formats do not produce globally unique identifiers. Another problem is that cataloging rules and formats exist in many variations around the world. Nor does traditional cataloging scale well – as the volume and variety of scholarly artifacts grows, the proportion of useful items that have full cataloging records decreases.

In the current environment, more than 15 years after the emergence of the World Wide Web, capabilities exist for globally persistent identifiers at the level of granularity necessary for scholarly objects. The Web Architecture allows any artifact available online in digital format – i.e., a Web resource – to be identified and accessed via a URI (Architecture of the World Wide Web, 2004; Uniform Resource Identifier, 2005). The Semantic Web introduces the notion of using URIs to identify non-document artifacts such as real-world objects (people, stars, cars, etc.) and concepts (the number zero) using URIs. The Linked Data instantiation of the Semantic Web also introduces an approach for describing such non-document artifacts (Bizer, Cyganiak & Heath, 2007).

Digital Object Identifiers (DOI) are routinely assigned to journal articles and also can be used to identify datasets (Digital Object Identifier System, 2009; Paskin, 2005). The motivations of publishers to assign DOIs include both long-term persistent identification and management of intellectual property. DOIs can be used outside of the Web context (they are incorporated in both paper and digital versions of articles, aiding in identification and retrieval). They can also be expressed as URIs using both the "HTTP" (Hypertext Transfer Protocol, 1999) and "info" ("info" URI Scheme, 2006) URI schemes. Hence, DOIs fit seamlessly in the general URI-based approach for the identification of scholarly artifacts.

Besides providing a framework for resource identification, the Web Architecture also allows properties and relationships between resources to be expressed. The standard widely adopted for this purpose is the *subject-predicate-object* model of RDF (Resource Description Framework, 2004). Associated specifications such as RDF/XML and Linked Data best practices (Bizer et al., 2007) detail how to serialize RDF-based descriptions into machine-readable formats and how to publish such descriptions to the Web. RDF-compliant vocabularies are emerging, both cross-community (e.g., Dublin Core Terms) and community-specific (Open Biomedical Ontologies, 2009), to express a wide range of properties and relationships.

These technical developments in Web services create the opportunity to operationalize the intellectual relationships between scholarly objects. Documenting relationships between objects was a cataloging luxury in a paper environment (Tillett, 1989; 2004). In the digital and distributed environment of eResearch, the ability to discover related resources becomes a necessity. Individual objects may have meaning only in relation to other objects. Furthermore, many artifacts in the eResearch realm are aggregations of others. A publication can be the composite of a digital manuscript, a dataset on which the findings reported in the manuscript are based, software used to derive findings from the dataset, and a video recording of experiments that led to the dataset. This list of components is by no means exhaustive. Examples of the many types of data that are of potential value for study or reuse include chemical compounds, astronomical observations, demographic surveys, and geographical maps. Until recently, these kinds of data were confined to hard drives, notebooks, laboratory cabinets, and refrigerators. Information sources that may have been discarded or left to deteriorate at the end of a project have become valuable objects in and of themselves (Borgman, 2007; Bowker, 2005).

These data – raw data, contextual data, semi-processed data, and so on – are important to scientific productivity and to scholarly communication. They may be the traces necessary to repeat an experiment, to reformulate a theory, to criticize a claim, or to make comparisons between places and over time. For example, sensor data are useless without related sensor calibration information, and the calibration information alone is meaningless. Lacking the legitimization and publishing process of traditional channels of scholarly communication, data often are lost irrevocably. In data-driven disciplines such as astronomy, seismology, and the ecological sciences, some observations are irreplaceable – comets, earthquakes, and spring blooms cannot be repeated for the sake of scientific pursuits. It follows that many scientific data are as important to knowledge as are the published papers that analyze and report them (Long-Lived Digital Data Collections, 2005; Cyberinfrastructure Vision for 21st Century Discovery, 2007; Borgman, 2007; Bourne, 2005; Lyon, 2007; Stodden, 2009a; b).

## RELATED WORK

The proliferation of individual artifacts and genres and the diffusion of responsibility for controlling them (authors, publishers, libraries, repositories, webmasters, etc.) makes urgent the need to find ways to link related digital objects. The relationships between objects are manifold, not all of which are useful for discovery. Identifying the relationships between artifacts deemed important by producers and potential users is essential. The forms and types of scholarly communication that result from a given research project is a rich area of study (Latour, 1988; Latour & Woolgar, 1986; Lynch & Woolgar, 1988). Scholars write grant proposals, working papers, conference papers, journal articles, and books. They give informal talks and keynote presentations. They use their data in their research and in their teaching. Little research has been done on the relationships between these types of scholarly objects, a body of work begun by Garvey and Griffith in the field of psychology (1964; 1966; 1967). Even less research exists on how to manifest the relationships between scholarly objects in ways that make them retrievable as related units (Van de Sompel, 2003). It remains difficult to determine what objects might be related to any given object, let alone to determine how they are related and how they vary. If the discovery of any object could provide an entry point into the set of related objects, information seeking and use would be improved substantially.

The scattering of scholarly content across many variants has led to a renewed interest in the intellectual relationships between the artifacts of a given research project. Frandsen and Wouters (2009), for example, have examined the processes by which working papers become journal articles. They focused their study in economics because it is one of the few fields in which such relationships are made explicit. Among their findings is that bibliographic references are both added and deleted in the transition from working paper to journal article, as the authors adapt the article for the target journal. The elapsed time between paper and article is a determinant of the degree of variation in references; the longer the time to journal publication, the more updating that occurs. While journal articles are usually shorter than the working papers on which they are based (Frandsen & Wouters, 2009), conference papers are typically expanded in length to become journal articles (Montesi & Owen, 2008). Journal articles themselves have many genres, such as theoretical, reviews, short communications, case studies, comment and opinion. These genres serve different communicative roles in individual disciplinary communities (Montesi & Owen, 2008). Scholars follow many different paths through genres and types of artifacts in assembling evidence for their writing, and these paths also vary considerably by discipline (Palmer, 2005). Complicating matters further, references to papers found in abstracting and indexing services such as the *Web of Knowledge* can vary considerably from references to Web-based sources of those same papers (Kousha & Thelwall, 2007). Academic blogs are becoming another source of links between scholarly artifacts, but they are even less consistent as a bibliographic mechanism (Luzón, 2009).

The scholarly communication process also has been conceptualized as a "value chain" (Borgman, 2007; Van de Sompel et al., 2006). The notion of value chain originated in the business community to describe the value-adding activities of an organization along the supply chain (Porter, 1985). Although the relationship between artifacts is rarely as linear as the term "chain" implies, scholarly processes often do follow a sequence of steps as, for example, manuscripts are revised and enriched en route to publication. A scholarly value chain might consist of the initial version of the manuscript (the preprint) submitted to a journal or conference, subsequent revisions of the manuscript, and supplemental material such as images, indexing terms, and the publisher's metadata.

Following the chain is difficult when these related items are distributed across the Web, as is often the case. For example, the published version of the manuscript and associated metadata might be available at the publisher's digital library site, the article preprint might be available in the author's institutional repository, and supplementary material might be found on the author's personal website. The linkage among these resources may not be explicit. Even if all these materials were from the same server, the linkage between them might only be discerned by human interpretation. A machine-readable representation of their scholarly relationships is missing (Van de Sompel, 2003). Research by Van de Sompel, Lagoze, Payette, and colleagues into this problem culminated in the release of the OAI-ORE specifications, introduced in the next section (Van de Sompel, Payette, Erickson, Lagoze & Warner, 2004; Warner et al., 2007).

## OPEN ARCHIVES INITIATIVE OBJECT REUSE AND EXCHANGE

The OAI-ORE (Object Reuse and Exchange, 2008) specifications – hereafter referred to as ORE – define a standard for the identification and description of clusters of Web resources (known as "aggregations"). ORE provides an aggregation with a URI, a description of its constituents, and optionally the relationships among them.

Let us suppose that a collection of resources related to a particular scholarly publication is available on the Web: a manuscript, its revision, a preprint version, and the published version with publisher's metadata and additional material. An HTML page at an institutional repository (commonly referred to as "splash page") might contain descriptive metadata (such as title, authors, publication venue) about the original manuscript, links to subsequent versions (e.g., the preprint and publication) and formats (e.g., in PDF and Microsoft Word) available from the institutional repository or elsewhere on the Web, and links to additional material (e.g., datasets and images related to the article) available on the Web. Each of these resources, including the splash page, is identified by a unique URI. When these URIs are "dereferenced" – that is, resolved via a common Web protocol (e.g., HTTP) – a human-readable representation of the resource is returned; it might be a manuscript in Microsoft Word format, a publication in PDF format, or the publisher's splash page for the publication (Lagoze & Van de Sompel, 2008).

The implicit linkage between objects on a Web page may be apparent to a human reader, who can easily discern the structure and content of their representations. However, machine agents and Web services may fail to interpret these materials as related resources. ORE provides a data model to assemble the URIs of these resources into an aggregation. Figure 1 depicts all the components of the conceptual ORE aggregation for a scholarly publication: resources, on the right, are aggregated by the Aggregation A, which is described by Resource Map ReM. The URI A is the identity for referencing the aggregation. When URI A is dereferenced (i.e., resolved) by a human agent (e.g., via a Web browser), a descriptive HTML page (the "splash" page) is returned that details the aggregation in a manner suitable for human consumption. When URI A is dereferenced by a machine agent (e.g., a Web crawler), the Resource Map ReM is returned. Resource Maps are machine-readable documents that describe ORE aggregations and can be expressed in a variety of formats including Atom XML, RDF/XML, and RDFa. Note that ORE allows a hierarchy of aggregations, such that one aggregation can refer to another. However, each aggregation must have its own Resource Map.

Moreover, the Resource Map can leverage RDF to express relationships and properties for the aggregation and its constituents. RDF subject-predicate-object statements (commonly known as "triples") can express the relationships among these resources. Figure 1 shows two kinds of RDF statements. The first one, with predicate *dcterms:hasVersion* (The Dublin Core Metadata Initiative Terms, 2009), asserts the relationship between different versions of the same item, requiring three RDF statements: between manuscript and revision, between revision and preprint, and between preprint and publication. The second RDF statement, with predicate *foo:hasBibliographicDescription*, asserts the relationship between the published version of an article (Publication) and its bibliographic description (Publisher Metadata).

Figure 1 also reflects the multiple agencies that can be responsible for components of an ORE aggregation. The agents responsible for the creation of the artifacts presented in the grey box, on the right side of Figure 1, may be authors (e.g., the original manuscript, the revision, the preprint, etc.) or publishers (e.g., copy edited versions, metadata). Using either automated or manual techniques, aggregations can be generated by institutions, repositories, or authors. These different levels of agency are best illustrated by examples. The JSTOR repository is generating nested aggregations that detail the journal-issue-article-pages topology of the JSTOR collection (JSTOR, 2009; Van de Sompel et al., 2009). In this case, the agent responsible for asserting the aggregations is JSTOR itself. Supported by appropriate tools, authors can create and publish their own aggregations. For example, Microsoft is adding tools to the Office Word Suite that will improve workflow for scholarly publishing, including the capability to create ORE aggregations (Fenner, 2008). Hunter and her colleagues also have demonstrated the feasibility of creating desktop tools to generate and publish aggregations with rich relationship information (Hunter & Cheung, 2007). With such tools in hand, authors become the agents to create aggregations, and can link related resources early in the writing process, thereby producing semantic publications (Shotton, Portwin, Klyne & Miles, 2009).

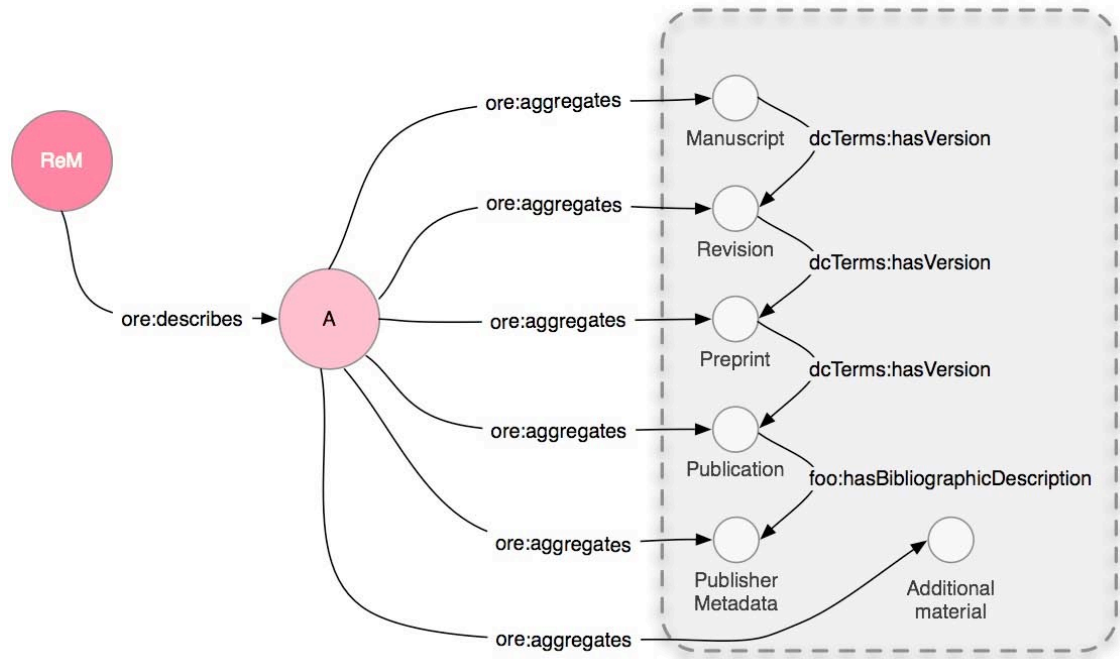

*Figure 1. ORE Aggregation representing a scholarly publication.*

The technical details of OAI-ORE, RDF, and Linked Data are available elsewhere (Resource Description Framework, 2004; Object Reuse and Exchange, 2008; Semantic Web Activity: W3C, 2009; Bizer et al., 2007). The succinct explanation of ORE presented above should suffice to demonstrate its benefits for the identification and description of scholarly artifacts that are aggregations of Web resources. The ORE approach introduces a minimal ontology aimed solely at addressing the resource aggregation problem. However, as is the case with all RDF-based approaches, ORE can be combined with other ontologies to achieve more expressive descriptions of the aggregated resources and their relationships. The true power of eResearch and the Semantic Web lies in representing the relationships that are most valuable for managing and discovering scholarly information. We argue that the value is best determined through study of scholarly and scientific practices. OAI-ORE data models are even more useful when combined with ontologies suited to individual communities of practice.

# RESEARCH SITE: THE CENTER FOR EMBEDDED NETWORKED SENSING

The research reported here is based in the Center for Embedded Networked Sensing (CENS), a National Science Foundation Science and Technology Center established in 2002 (Center for Embedded Networked Sensing, 2009). CENS supports multi-disciplinary collaborations among faculty, students, and staff of five partner universities across disciplines ranging from computer science to biology, with additional partners in arts, architecture, and public health. More than 300 students, faculty, and research staff are associated with CENS. The Center's goals are to develop and implement wireless sensing systems and to apply this technology to address questions in four scientific areas: habitat ecology, marine microbiology, environmental contaminant transport, and seismology. CENS also has projects concerned with social science issues, ethics and privacy, and citizen science.

Our research on scientific practices addresses questions about the nature of CENS data and about how they are produced and managed. We have constructed tools and services to assist in scientific data collection and analysis (Mayernik, Wallis & Borgman, 2007; Pepe, Borgman, Wallis & Mayernik, 2007) and have used CENS data to teach middle school and high school science (Borgman, Wallis & Enyedy, 2006; 2007; Wallis, Milojevic, Borgman & Sandoval, 2006). Particularly relevant to the work presented here are our past efforts to pursue ways to represent the scientific life cycle of CENS research (Borgman, Wallis, Mayernik & Pepe, 2007) .

## The Life Cycle of Embedded Networked Sensor Research

The notion of a "life cycle" can refer to human activities or to information. When referring to human activities, as in science, a life cycle encompasses the stages and pursuits of a particular field of practice. In reference to information, a life cycle embodies the changing status of an information object over its "lifetime." Documents (or other information objects) may originate for one purpose, e.g., to describe an experiment, and be sought for other purposes later, e.g., as legal evidence for priority of discovery. They may be in active use for some time and then lay dormant or be discarded. The information life cycle is a fundamental concept in archives, documentation, and library science (Gilliland-Swetland, 2000).

Our research encompasses both kinds of life cycles. We consider the scientific life cycle to be *the socio-technical ensemble of activities of a particular field of practice and the associated artifacts*. The scientific life cycles discussed in this article involve activities, workflows, stages and products specific to the field of embedded networked sensing with application in the environmental sciences (Estrin, Michener & Bonito, 2003; Michener & Brunt, 2000; Szewczyk et al., 2004) and seismology (Ahern, 2000; Suarez et al., 2008). We have found that life cycles, practices, and products vary between individuals and research teams. Environmental field work, for example, can be characterized by whether researchers identify a research problem in the field or in the laboratory, how they locate field sites in which to test or generate hypotheses, how they assess field sites for appropriate positioning of data collection equipment and sample acquisition, and the ways in which they calibrate equipment in the laboratory and the field (Borgman, Wallis & Enyedy, 2007). Later studies refined our understanding of these research activities, revealing a scientific life cycle that begins with the design of an experiment, followed by the calibration of the sensors, the capture and cleaning of the data, its analysis and the publication of the experimental results (Wallis, Borgman, Mayernik & Pepe, 2008).

The scientific life cycle that we derived from these studies in the context of environmental sensing research is presented in Figure 2. The order and position of the stages depicted in the image are not absolute. Some stages are iterative and some may occur in parallel. In certain cases, some stages might be skipped altogether (e.g., the calibration of instruments or the preservation of data). The timescale of the life cycle depends on the context of research. This type of embedded networked sensing research is often conducted as a "campaign," which may last from a few hours to a week or more in length.

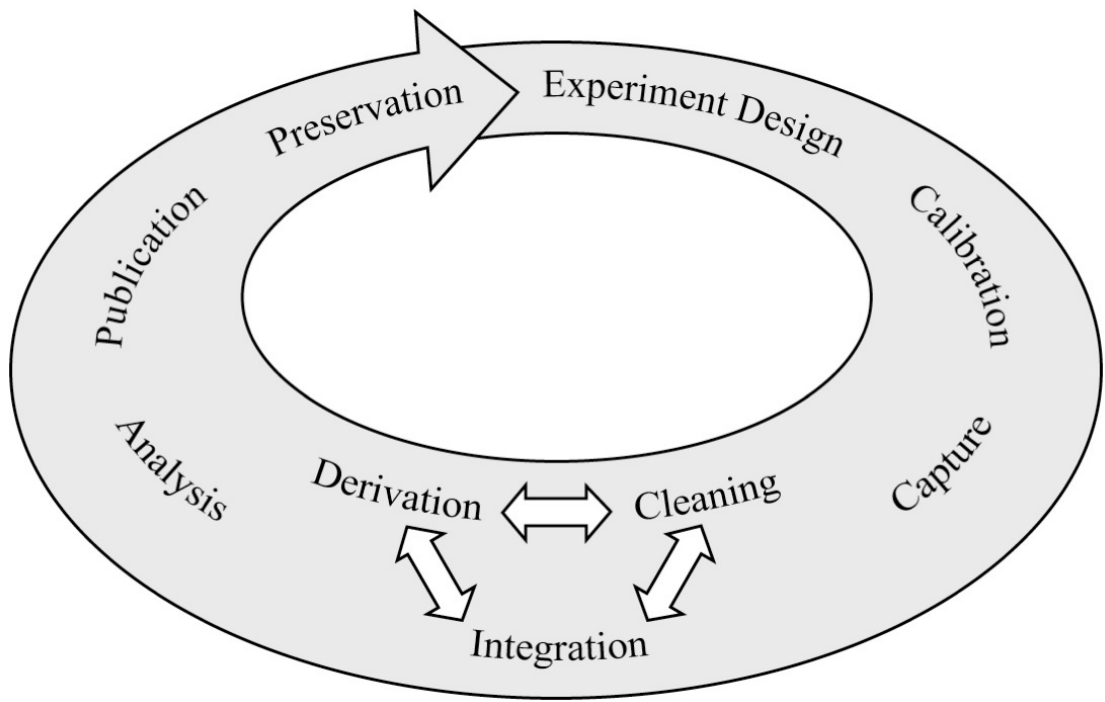

*Figure 2. The scientific life cycle in the context of environmental sensing research (Wallis et al., 2008).*

**The Integrated Life Cycle: Connecting Practices and Products**

Figure 2 represents the life cycle of scientific practices in environmental sensing research. Implicit in the figure are the products that may result from – or be necessary for – certain practices. In the experimental design stage, for example, laboratory notes and deployment plans are produced. The calibration stage can include equipment calibrations both in the laboratory and in the field, as equipment often is recalibrated to reflect field conditions. Artifacts such as lists of equipment taken into the field and the condition of that equipment may be produced at the planning stage or may be documented more fully during and after data collection. In the data capture phase, records on the initial placement of sensors, movement of sensors, and decisions made in the field may be produced. This array of contextual information about a field study can be essential documentation for interpreting results and for planning subsequent field research. To account for this set of scientific artifacts, Figure 3 integrates the life cycle of environmental sensing research (of Figure 2) with the larger range of scientific products identified in our study of this scientific community.

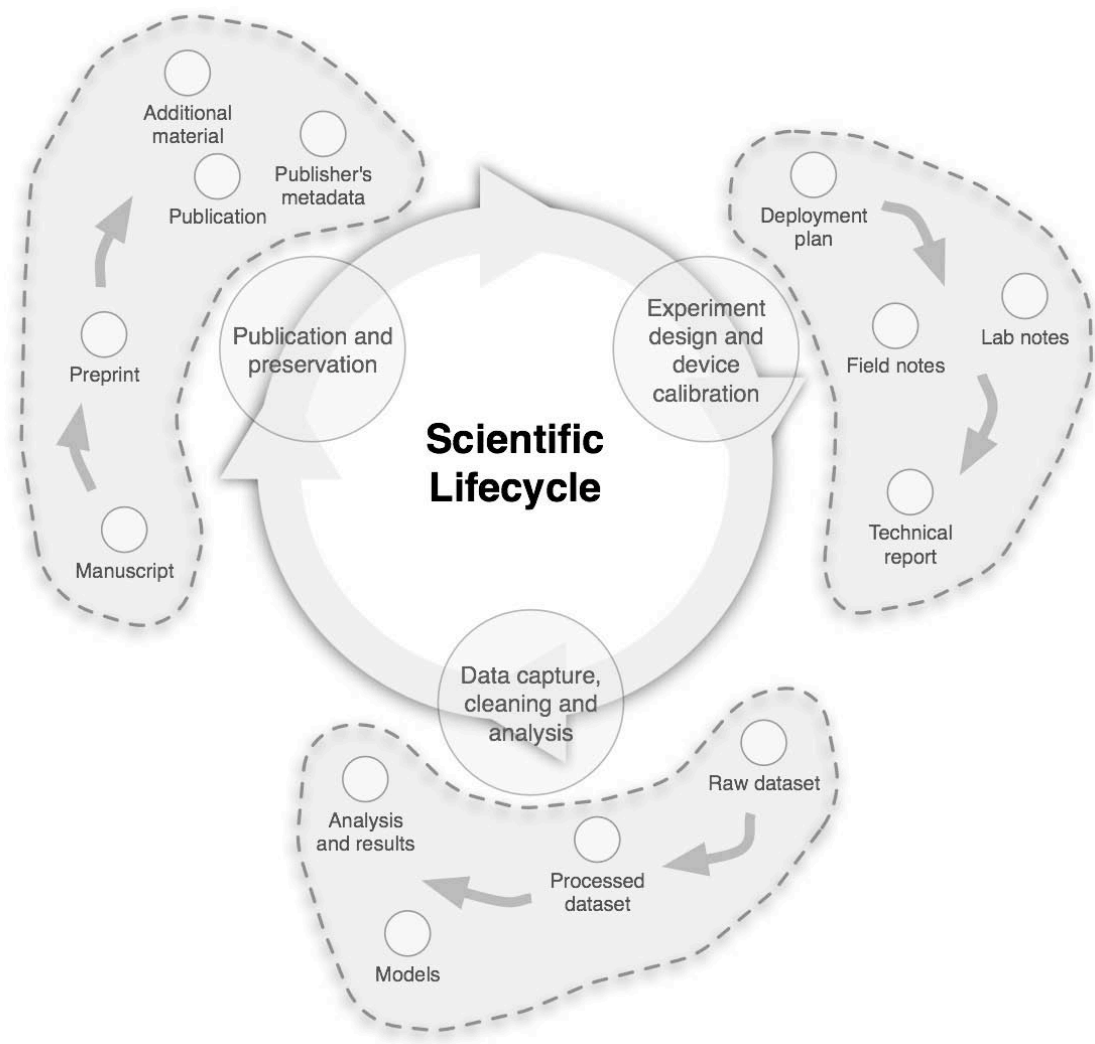

*Figure 3. The integrated scientific life cycle of embedded networked sensor research.*

The inner circle in Figure 3 represents the life cycle of scientific research in environmental sensing. It is a modified version of Figure 2: the steps of the life cycle have been condensed into three stages: (i) experiment design and device calibration, (ii) data capture, cleaning and analysis, and (iii) publication and preservation. This categorization reflects our understanding of the CENS data ecology, based on ethnographic observations and interviews with participants. We performed a large-scale exploration of the CENS data ecology to classify both the scientific practices through which artifacts are produced, handled, and exchanged by CENS researchers, and the artifacts themselves. The results of this study, presented in detail elsewhere (Pepe et al., 2007; Wallis et al., 2008) confirmed the occurrence of these three life cycle stages characterized by their data types.

Clockwise, from the top of the inner circle, the life cycle begins with the design of new experiments. Initial experiment design typically involves sketching out a deployment plan, in which the details of the proposed research are described. Deployment plans, depicted in the outer circle, are often the earliest artifacts produced in the scientific data life cycle. The life cycle proceeds with the calibration of the sensor devices to identify the offset between actual measurements and expected measurements. Calibration procedures are described in lab and field notebooks and may also be included in technical reports.

The next stage in the life cycle is data capture, which consists of observation and monitoring of specific phenomena in the field and results in the collection of raw data. The capture phase may produce multiple types of data. In recent field work, we identified four categories of data that resulted from a single "campaign" of field research in environmental sensing: sensor-collected application data, hand-collected application data, sensor-collected performance data, and sensor-collected proprioceptive data. The first two categories are of most interest to the application scientists (usually biologists in this example) and the latter two categories are of most interest to the computer science and engineering researchers. Each of these groups may maintain their own data separately. Each of these datasets may be converted to another format, augmented, filtered, or analyzed in a number of ways, resulting in different versions or "states" of the same dataset. Interpreting the results of the project, not to mention reconstructing the field campaign, may require access to many or all of these datasets (Borgman, Wallis, Mayernik et al., 2007). After capture, raw datasets are then "cleaned up" to account for inconsistencies such as calibration offsets and statistical errors. Refined data usually are analyzed statistically.

The life cycle ends with the scholarly publication process: preparing manuscripts, submitting them to conferences or journals, and revising them until publication. With publication, a number of other artifacts are also preserved, such as supplemental materials, citations, images, and publisher's metadata.

It is worth reiterating that the integrated life cycle presented in Figure 3 is based on the social, cultural, academic practices, and workflows of a specific scientific domain: embedded networked sensing research. Clearly, life cycles will vary widely by type of scientific practice, from laboratory to field, by research methods, and by research questions. Yet, for the purposes of information discovery in this specific distributed environment, we have identified three major life cycle stages that result in artifacts such as datasets, laboratory notebooks, and publications. We argue that even though these artifacts may be useful individually, their value is greater if they are linked together to form an integrated scientific life cycle. In the remainder of this article, we demonstrate the use of the ORE data model to represent the integrated scientific life cycle of two applications that employ embedded networked sensing technologies.

## CASE STUDIES IN EMBEDDED NETWORKED SENSING RESEARCH

We demonstrate how ORE can be applied to embedded sensing research by offering two contrasting case studies: a seismological project and an environmental project of CENS (Center for Embedded Networked Sensing, 2009). We are studying both of these projects as part of ongoing research on data practices. Our studies examine how CENS' researchers collect, organize, manage, use, share, and archive data (Borgman, Wallis, Mayernik et al., 2007; Wallis et al., 2008). We also design and build systems to facilitate the collection and sharing of the center's data and metadata (Mayernik et al., 2007; Pepe et al., 2007). In the following cases, we illustrate the diversity of information products that are created during embedded sensor research and the diversity of current methods for storing, sharing, and making these products available. This is a conceptual demonstration of the use of ORE to connect these products in ways that reflect scholarly and scientific practices. We address the potential of ORE for information management and retrieval, while identifying challenges to its implementation in research environments.

The two case studies presented here are based on ethnographic research of CENS field deployments and on interview and ethnographic studies conducted earlier, as discussed above. To frame the following discussion, we note that a sensing system "deployment" is a research activity in which sensors, sensor delivery platforms, or wireless communication systems are taken out into the field to study phenomena of scientific interest. CENS deployments have taken place at numerous locations around the world, including Bangladesh, Central and South America, and California lakes, streams, and mountains. Members of our research team have studied thirteen CENS sensor system deployments as participant observers, both observing and taking part in deployment activities, encompassing approximately forty days of participant observation over two years. The observed deployments span six CENS projects, and the number of CENS researchers participating in the deployments ranged from two to ten. Ethnographic field notes and digital photographs focused on the nature of deployments, field-based scientific research practices, and the role of information systems in this heavily instrumented field-based research. Our participant observations have been supplemented by informal interviews and discussions before, during, and after deployments with CENS researchers regarding their data collection and collaboration practices, as well as regular interaction with CENS researchers at formal center gatherings, such as research reviews and retreats, weekly research seminars, and informal gatherings and discussions in labs and offices.

# Life Cycle of a Seismology Research Project

Seismology is the "grandfather application" of embedded networked sensing research. This community has the longest history of using sensor technology of any of the CENS partners. Seismic equipment is more robust, but also much more expensive and less portable than most of the sensor technologies used in other CENS scientific applications. The scientific and technical expertise of the seismic community has been central to CENS since its inception.

While some of the seismic applications are local and short in duration, they also have the largest scale projects of CENS and some of the longest in duration. In the MASE project (Middle American Subduction Experiment, 2009), CENS researchers deployed 50 radio-linked and 50 standalone seismic stations across Mexico for approximately two years, from 2006 to 2007. When the MASE project was completed, the seismic sensing stations were moved to Southern Peru, where they were installed in 2008 for an expected two-year duration. These joint UCLA and Caltech projects installed seismic sensors and wireless communication equipment at approximately 5km intervals from the Pacific Ocean into the continents, a range of several hundred kilometers. CENS communication systems enabled the seismic researchers to transport their data wirelessly from the sensor installations back to UCLA via the Internet. For the seismic projects, the sensing equipment is left to capture data autonomously for months at a time, thus security and robustness of equipment are of great concern.

The first stages of the life cycle of these seismic sensing deployments occur long before any seismic data are collected. Products associated with this first stage of the life cycle contain planning, provenance, and contextual information about a deployment. Initially, researchers identify suitable locations and contact relevant landowners or administrators for permission to install equipment. Documents created at this stage include interim reports that describe the practical details of the experiment, such as a deployment plan, digital and paper maps of land and of seismic topography, letters of permission, payment agreements, and other documentation describing the research sites. These deployment records contain rich information about the social and technical practices of these research projects, and are of great value both for interpreting research results and for planning future projects.

Once sites are selected, sensing and communication equipment is installed. At each site, researchers dig holes and pour concrete, place and orient the seismic sensor, install electronics, radios, masts, and antennas, and wire the entire station with power, either from a nearby building or from a solar panel they install themselves. Radio electronics at each site must be configured with the most current version of the wireless transmission protocol. Details about the equipment installation, including problems encountered in the field and how they were addressed are recorded on field notes and shared among the team through emails. This information about field activities is also stored in the CENS Deployment Center (CENSDC), a database of deployment information developed by our data practices team (Borgman, Wallis, Mayernik et al., 2007; Mayernik et al., 2007; Wallis et al., 2008; Wallis et al., 2007). Once recorded in the CENSDC, researchers have the option to make the products publicly available. The CENSDC maintains context information about the deployment that is discoverable by current and future research teams. In turn, this context information supports a much richer interpretation of the datasets and publications to which they are linked.

ORE can be used to represent the conceptual linkage that exists among artifacts discussed thus far, and associated with this first stage of the scientific life cycle, as displayed in Aggregation A-1 of Figure 4.

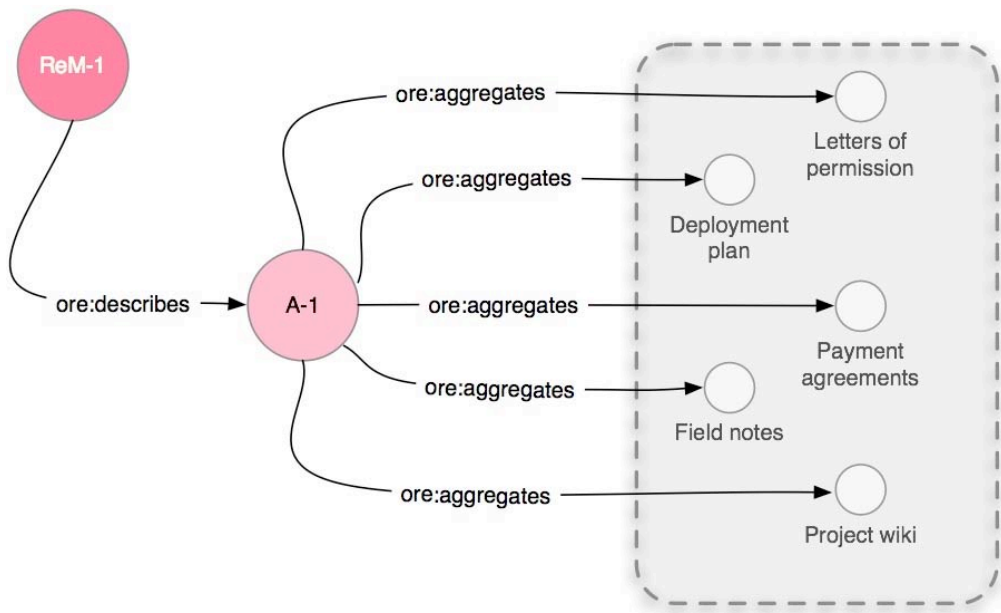

*Figure 4. ORE Aggregation representing the first stage of the scientific life cycle of a sensor network application in seismology (experiment and deployment planning).*

After the sensors are installed and functional, seismic data are collected. Initially, data are stored on a compact flash card within the site electronics. If the wireless communication systems are working, the data are then transmitted to Internet hubs and back to UCLA using CENS point-to-point wireless routing protocols. If the wireless communication systems are not working, data are stored on the flash card until manually downloaded by a member of the research team or until wireless communication is restored. Data are initially stored and transmitted as files in *Mini-SEED* format (Standard for the Exchange of Earthquake Data, 2009), a binary format that specifies a standardized structure for the collection of seismic data. At UCLA, the data are converted into the SAC (Seismic Analysis Code, 2009) data file format, another binary file format for time-series seismic data, and then are sent to Caltech for inclusion in the main project database. The pre-converted data are kept at UCLA in a local database. The converted data are held at Caltech for long-term storage. All subsequent analyses use the converted SAC files. Technical details for the wireless data routing protocols and the data conversion process are shared among the team on project wiki pages.

The seismic research team collects considerable amounts of contextual information about their projects. Project servers and websites are used to share pictures of installations, maps, and computer code, among other things. Additionally, for the Peru project, the group collects regular status updates regarding the health of the wireless network. These status updates consist of information such as radio battery life, amounts of data collected, available space on the flash card, times of system re-boots, and an assortment of other technical characteristics of the network. The updates are autonomously collected at each site on an hourly basis and transmitted with the seismic data from each site, but are stored separately from the converted SAC data files. Instead, they are stored on a project database and made accessible to the team on a group Web site. Not all of these data are stored in publicly accessible servers. Some are available only on password-protected servers for security reasons, including any data that contains GPS locations of currently installed equipment, such as the contextual data uploaded onto the CENSDC.

In sum, the data collection stage of the life cycle results in the following products, as displayed in the Aggregation A-2 of Figure 5: the collected seismic data, potentially available in two formats (Mini-SEED and SAC), and represented by Aggregation AR-2; deployment contextual data; and technical data relative to the health of the network. For diagram clarity, the Resource Map of Aggregation AR-2 has been omitted from Figure 5.

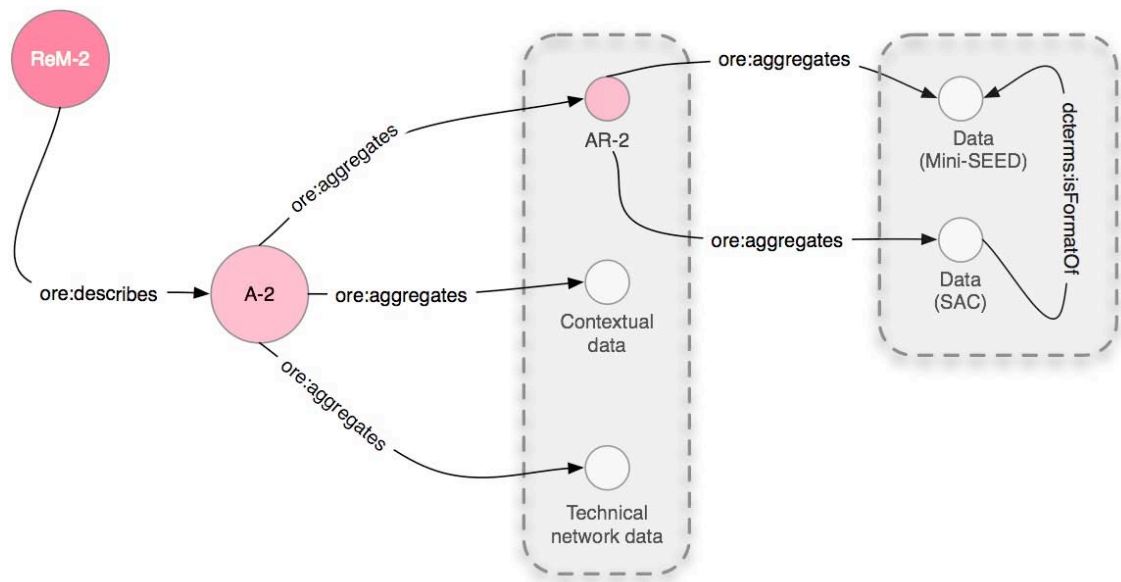

*Figure 5. ORE Aggregation representing the second stage of the scientific life cycle of a sensor network application in seismology (data collection).*

Publications from seismic projects span scientific and technical aspects of their experiments. The first publications from the MASE project addressed the computer science theory of the wireless communication protocols (Lukac, Girod & Estrin, 2006; Lukac, Naik, Stubailo, Husker & Estrin, 2007) and technical requirements for the experiment (Husker et al., 2008). Publications about the geophysical data per se were produced later in the project timeline (Husker & Davis, 2009; Perez-Campos et al., 2008; Song et al., 2009), with additional publications still in progress. Thus the life cycles of the computer science research and the geophysical research operate on different timelines, despite collaboration on a single project. Two of the three technical publications that have appeared to date are available online in two different places: as pre-prints in the CENS section of the University of California's institutional repository (CENS eScholarship Repository, 2009), and as publications on the publisher's websites. The other one (Lukac et al., 2007) was published as a technical report, and is available only in the CENS institutional repository. These three publications and their associated pre-prints can be collected into an Aggregation, A-3, as displayed in Figure 6.

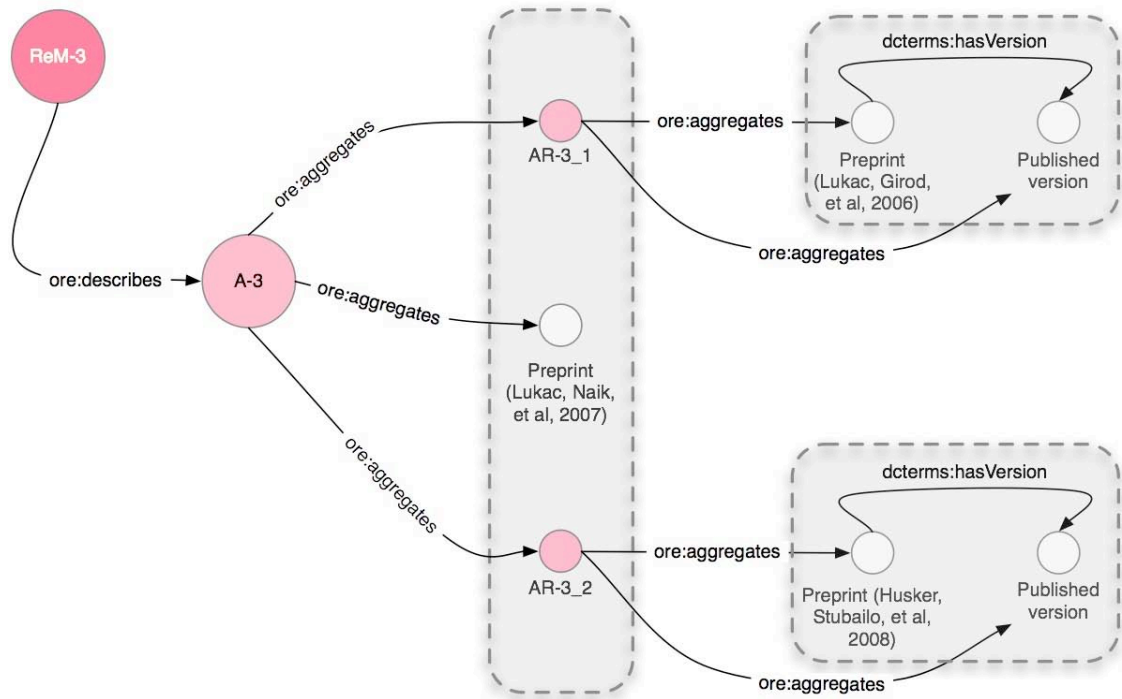

*Figure 6. ORE Aggregation representing the third stage of the scientific life cycle of a sensor network application in seismology (publication).*

The three ORE Aggregations (A-1, A-2 and A-3) from Figures 4, 5 and 6 can be linked to reconstruct the scientific life cycle of this seismic project, as displayed in Figure 7. A Resource Map, ReM-t, describing the entire scientific life cycle as an aggregation with citable identity A-t can be published to the Web, where it can be harvested and leveraged by other applications. When A-t is dereferenced it will return two descriptions: Resource Map ReM-t, suitable for machine agents, and an HTML splash page, suitable for humans. Both descriptions reveal the components of the scientific life cycle and any relationships among those components. In Figure 7 some ORE and RDF relationships and all the Resource Maps have been omitted for diagram clarity.

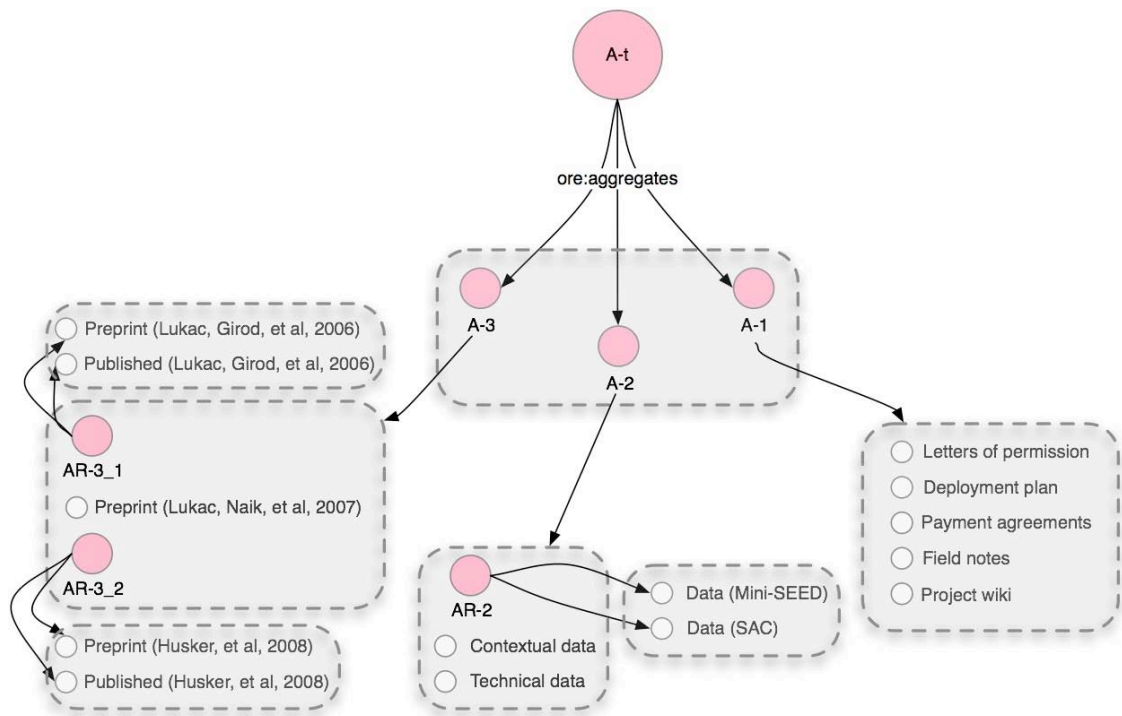

*Figure 7. ORE Aggregation representing the entire scientific life cycle of a sensor network application in seismology.*

**Life Cycle of an Environmental Science Research Project**

Environmental science is another fertile application of CENS technologies. A team of electrical engineers and computer scientists from UCLA collaborates with a team of environmental scientists and engineers from the University of California, Merced, to develop sensing technology for aquatic systems. The environmental science group has used CENS Networked Info-Mechanical Systems (NIMS) technology to conduct research on the transport of various contaminants in the San Joaquin River basin in central California (NIMS: Networked Infomechanical Systems, 2006).

The core of NIMS technology is a mobile robotic platform that enables scientists to move sensing equipment through an environment in a precisely controlled fashion. To study the way contaminants move through a watershed, researchers hang a cable system across the rivers being studied and manipulate a NIMS unit along the cable. Sensors are attached to the NIMS platform and lowered into the water. They are then moved across the river by the NIMS machinery, with the system stopping at regular intervals to lower the sensors vertically through the water, enabling the scientists to create two-dimensional profiles of contaminant flow downstream through the transect. Whereas the seismic projects described above install sensing equipment that will be left to capture data autonomously for months at a time, these environmental sensing projects install sensor networks for short campaigns of data collection, with human observers in attendance. They tend to use "research grade" equipment that requires adjustment in the field; design and evaluation of the technology is usually part of the research.

Of particular interest is a project that has conducted at least three campaigns with NIMS technology since 2005 near the confluence of the San Joaquin and Merced rivers in Central California. In addition to deploying the NIMS system in the manner described above, numerous other sensors are used to collect data on water flow in and out of the river system. The first stage in the scientific life cycle of a typical NIMS campaign is planning, followed by the calibration of the sensors. These calibration metrics are documented in multiple technical reports. The UC-Merced team created a digital library to maintain records on these projects, both for internal group use and for public use on the Web (Sierra Nevada San Joaquin Hydrologic Observatory, 2009). Their digital library has sets of nested files, containing a wide variety of information on these contaminant transport projects. For example, technical reports documenting calibration and preparation of the NIMS device are listed in the *Sensors and Loggers* category. Documentation of the initial set-up of devices for specific deployments is listed in the category *Field Work*. Another category lists the software to control the NIMS unit. Some of the documentation about deployments, such as lists of equipment and team members, is also recorded in the CENS Deployment Center (CENSDC) system described above. The artifacts produced in the planning, experiment design, and device calibration stages of the life cycle for these environmental sensing projects can be grouped in an ORE aggregation. Figure 8 depicts this aggregation, consisting of four information resources that are physically located in two disjoint data libraries.

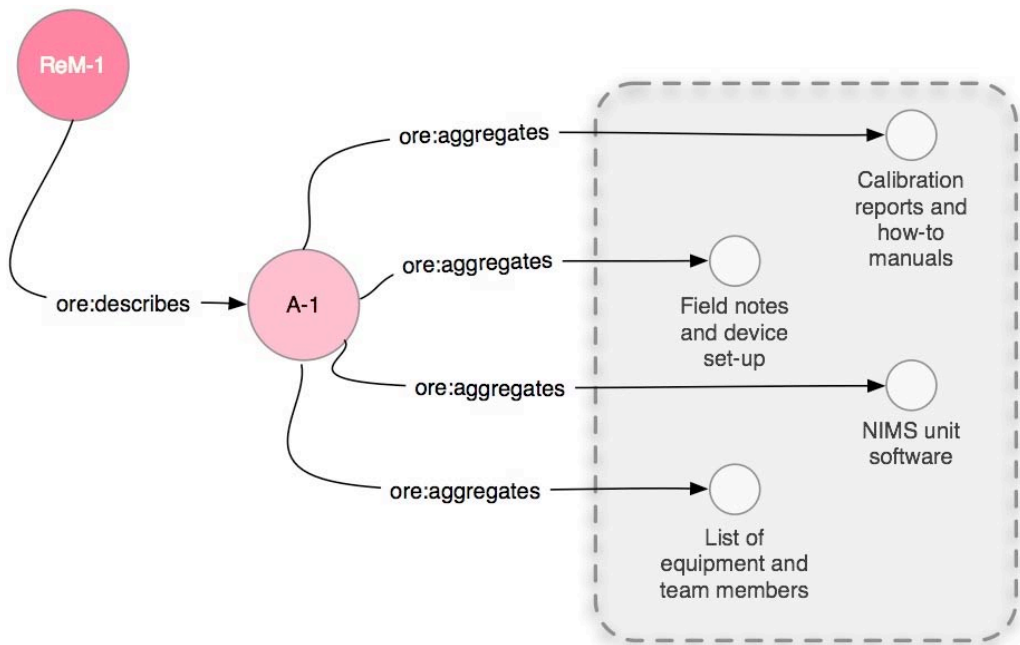

*Figure 8. ORE Aggregation representing the first stage of the scientific life cycle of a sensor network application in environmental science (design and device calibration).*

The next stage of the scientific life cycle for the NIMS campaigns begins with data capture. The sensors attached to the NIMS machine primarily capture data that relate to the water contamination levels. Other background data are collected as well, including ambient temperature, humidity, and barometric pressure from a mobile weather station, and bathymetry data to map the physical contours of the river banks and bottom. Contextual data are created, such as photos of the NIMS deployment site and related media files. These resources, along with the software used to perform post-processing and data analysis, are all publicly available through the UC-Merced digital library (Sierra Nevada San Joaquin Hydrologic Observatory, 2009).

The resources of this stage of the life cycle can be grouped together into an ORE aggregation representing the data capture and analysis stages of the life cycle (Figure 9). Raw data are often available in multiple formats (txt, csv, kml, etc.). For reasons of diagram clarity, the RDF relationships between these related data types are not shown in this figure.

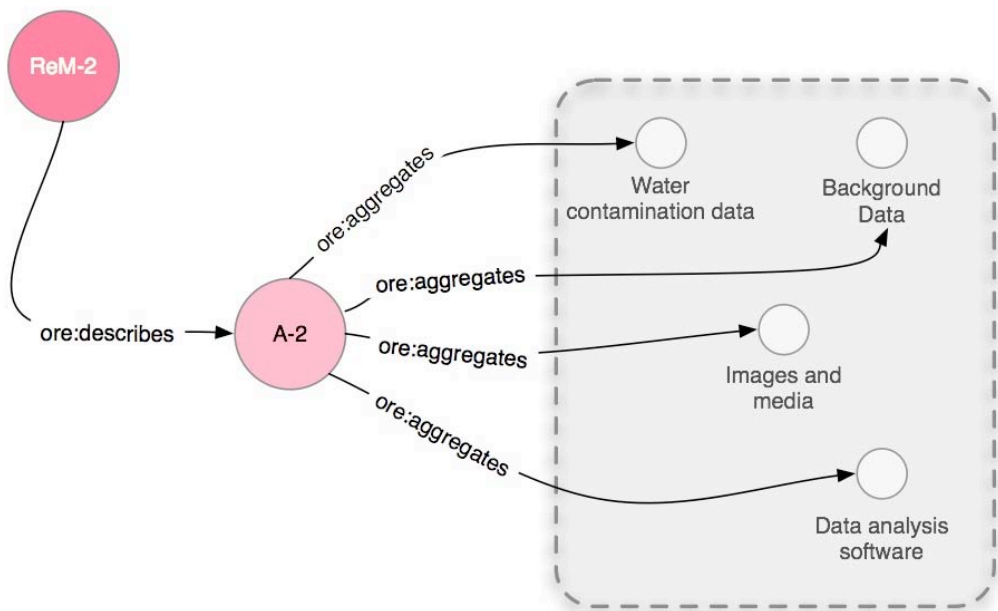

*Figure 9. ORE Aggregation representing the second stage of the scientific life cycle of a sensor network application in environmental science (data capture and analysis).*

As with the seismological sensing project presented in the previous case study, publications from environmental sensing span science and technology. The technical development of NIMS is documented in numerous technical reports, conference papers, journal articles, and Master's theses. The scientific aspects of using NIMS to study environmental phenomena in water contaminant transport are reported in multiple conference papers. Papers from this project are distributed across several archives, as are those from the seismic project. For example, a technical article (Singh et al., 2008) and a scientific article (Harmon et al., 2007) are available both as preprints at the CENS institutional repository (CENS eScholarship Repository, 2009) and as published articles in digital libraries of their respective publishers, i.e., the journal *Environmental Engineering Science* and SpringerLink, the publisher archive for the book series *Field and Service Robotics*. These scientific resources, available through two different archives, can be grouped together in an aggregation representing the final stage of the life cycle: publication and preservation (Figure 10).

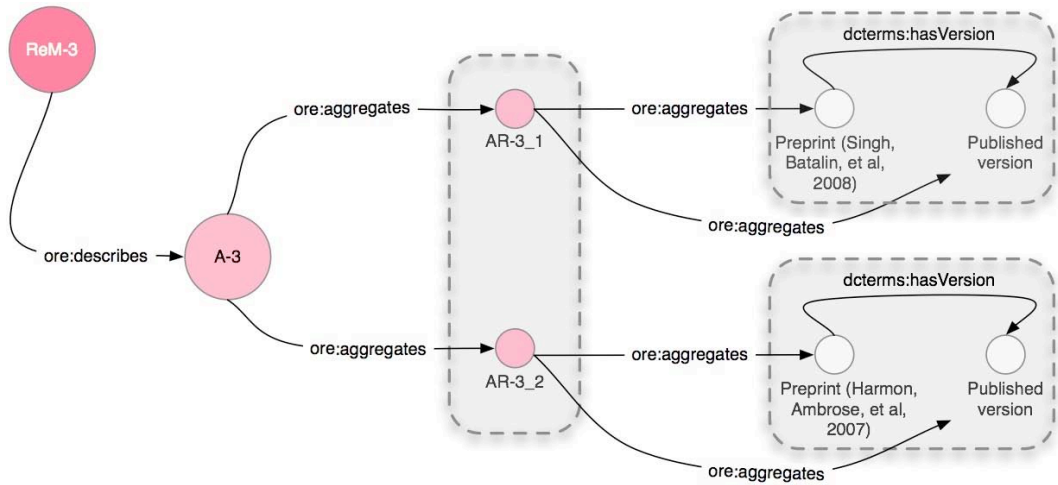

*Figure 10. ORE Aggregation representing the third stage of the scientific life cycle of a sensor network application in environmental science (publication and preservation).*

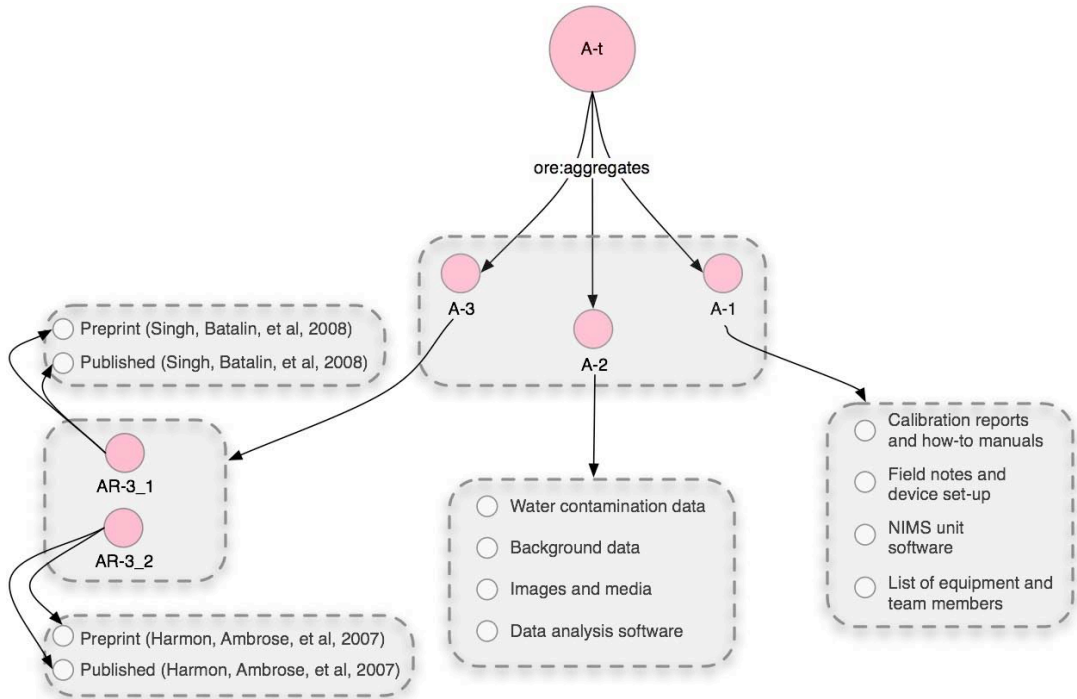

*Figure 11. ORE Aggregation representing the entire scientific life cycle of a sensor network application in environmental science.*

Similar to the previous case study, the aggregations generated in this application (in Figures 8, 9, and 10) can then be linked together (Aggregation A-t) to reconstruct the life cycle of this environmental sensing research project, as shown in Figure 11.

## DISCUSSION AND CONCLUSIONS

The ability to discover and retrieve related digital artifacts has not kept pace with the proliferation of digital scholarly resources. As eResearch – data- and information-intensive, distributed, collaborative, and multidisciplinary research – becomes the norm, the need for networked approaches to representing scholarly information objects becomes ever more urgent. Cataloging approaches are useful, but far from sufficient, as they lack the unique and persistent identifiers and the relationship expressions necessary for the management and discovery of distributed resources. The components of an architecture to aggregate scholarly resources are now in place, after many years of research in information organization and many technical advances in Web services. What is needed now is some rational basis on which to implement that architecture. The Open Archives Initiative's Object Reuse and Exchange data model would be of limited value if used to represent arbitrary relationships between resources. If the choice of relationships is based on scholarly and scientific practices, then the resulting aggregations should aid researchers in managing their own resources, assist information professionals in organizing resources, and help present and future users of those resources to discover, retrieve, and use them.

If Resource Maps describing scientific life cycles were commonly published to the Web, within or across the boundaries of scientific disciplines, Web harvesters could gather these descriptions and merge them into a database. As scientific assets are reused by the same and other scholars, multiple Resource Maps will reference the same artifacts. The resulting union of Resource Maps will form a rich graph that interconnects scientific artifacts. That network will reflect relationships among scholarly practices. Such a graph can be exploited by a variety of applications, including search, analysis, and visualization. A person – or a machine agent – could enter the graph at any point and trace relationships across time and place, offering unprecedented insights into the dynamics of scholarship.

Matching a technical architecture to practice is no small task. We have participated in the development of the ORE data model using the CENS research center as a testbed to assess its capabilities. At present, the ORE specifications are in their first production release (v. 1.0) and a number of library and repository services are in the process of implementing them in their systems to support the exchange of scholarly resources on the Web. More tools and mechanisms to facilitate the creation of ORE aggregations will become available through this implementation process.

The work presented here is a conceptual implementation of ORE to cluster data, documents, and an array of contextual information generated throughout the life cycle of scientific research in seismology and environmental applications that deploy embedded networked sensing systems. The CENS case studies presented here reflect the complexity of scholarly and scientific practices that need to be represented. Our initial experiments, working in concert with the ORE development process, indicate that ORE offers a feasible and promising approach to information management. At our current stage of work, creating aggregations of resources is a manual process. As we come to understand better the conceptual relationships between related resources, we hope to automate more of the process, connecting related scientific artifacts incrementally and automatically.

The next stage of our research at CENS is to implement ORE for discovery across the three digital libraries that contain publications, data, and deployment records. This approach will serve a larger goal, which is to capture data as cleanly as possible and as early as possible in the scientific life cycle. Once the integrated platform is operational, we plan to evaluate its effectiveness in practice. We will study how easily aggregations can be produced, either manually or automatically, and how useful they are both to CENS researchers and to information seekers. These evaluations will offer guidance to other information researchers on the potential value and limitations of this approach to improving the management, discovery, and retrieval of scholarly objects in a networked world.

## ACKNOWLEDGEMENTS

We thank David Fearon, Andrew Lau, Katie Shilton, and Jillian Wallis of the Department of Information Studies at UCLA for reading and providing comments on earlier versions of this article. Paul Davis, Professor of Earth and Space Sciences, UCLA, and Thomas C. Harmon, Professor of Engineering, University of California, Merced, kindly gave permission to present research conducted under their direction in our case studies in this paper.

CENS is funded by National Science Foundation Cooperative Agreement #CCR-0120778, Deborah L. Estrin, UCLA, Principal Investigator; Christine L. Borgman is a founding co-Principal Investigator and research lead of the Data Practices Team. The CENS Deployment Center is supported by core CENS funding. Alberto Pepe and Matthew Mayernik's participation in this research is supported by a gift from the Microsoft Research Technical Computing Initiative.